\documentclass[twocolumn]{apex}

\title{Metallic Pattern Fabrication in Organic Mott Insulating Crystal 
by Local X-Ray Irradiation}

\author{Naoki Yoneyama$^1$, Takahiko Sasaki$^1$, Norio Kobayashi$^1$,
Yuka Ikemoto$^2$, Taro Moriwaki$^2$, and Hiroaki Kimura$^2$}
\inst{
$^1$Institute for Materials Research, Tohoku University, 2-1-1 
Katahira, Aoba-ku, Sendai, Miyagi 980-8577, Japan

$^2$Japan Synchrotron Radiation Research Institute (SPring-8), 1-1-1
Kouto, Sayo-cho, Sayo-gun, Hyogo 679-5198, Japan}

\abst{
We have fabricated a metallic structure in an organic Mott insulator
$\kappa$-(BEDT-TTF)$_2$Cu[N(CN)$_2$]Cl.
The periodic metallic domains of approximately 90$\times$90 $\mu$m$^2$
obtained by X-ray irradiation through a molybdenum mesh mask
are visualized by scanning microregion infrared reflectance 
spectroscopy technique.
No deterioration by irradiation is found in a range of nanometer to 
micrometer scales.
We demonstrate that the present processing method is applicable
for the fabrication of molecular electronic devices.
}

\begin{document}
\maketitle

Organic materials have been of great interest for recent years because of its
potential for the application to electronic devices such as
light-emitting diodes, field effect transistor (FET), etc.\cite{OFET}
Their low-cost and lightweight feature is an important advantage over 
conventional semiconductors.
However, there is a difficult problem to utilize organics for fabricating devices:
softness of organics constantly interferes with assembling circuits,
particularly the preparation of good electrical contact between metal electrodes and 
organic surface.

The most widely used molecular semiconductors, including
pentacene, thiophene oligomers, and rubrene, etc., are conventional band insulators. 
For further performance improvement in organic electronics, Mott 
insulators have been the object of recent studies as a new seeding material.
Molecular based charge transfer (CT) salts stand out as a good candidate 
for this purpose, because most of them are included in 
strongly correlated electron systems, exhibiting a wide variety of 
electronic states by controlling bandwidth or carrier density. 
For instance, an organic FET system using 
tetrathiafulvalene-tetracyanoquinodimethane (TTF-TCNQ) based Mott 
insulating crystals has been successfully realized\cite{OFET3}.
One of the most vigorously investigated CT salts is so-called $\kappa$-type
bis(ethylenedithio)tetrathiafulvalene (BEDT-TTF) family, which has 
various electronic states resulting from the strongly correlated 
electron interaction \cite{1106}.
The ground state of an antiferromagnetic insulator 
$\kappa$-(BEDT-TTF)$_2$Cu[N(CN)$_2$]Cl varies to a superconducting state by 
applying small static pressure ($\sim$ 300 bar) \cite{235,327,986,971}.
This has been known to be a Mott insulator-to-metal transition induced 
by controlling bandwidth.
On the other hand, carrier density is almost invariable in the CT salts 
because of difficulty in substituting molecules partly with different 
valent components\cite{BF1} or synthesizing nonstoichiometric crystals except for
a few cases\cite{BF2,BF3}.

\begin{figure}
\begin{center}
\includegraphics[clip,width=8cm]{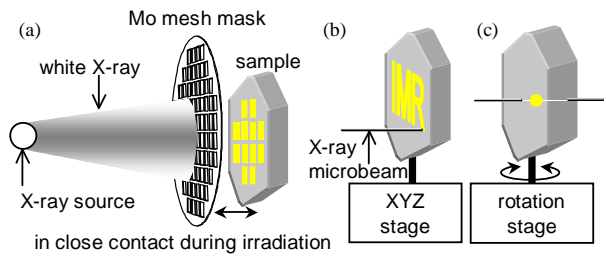}
\end{center}
\caption{
Schematic view of (a) the configuration of X-ray irradiation to the sample
through molybdenum mesh mask,
(b) the application for an electrical circuit drawing by using X-ray microbeam,
and (c) for an metallic dot by X-ray convergent projection.
}
\label{f_schematic}
\end{figure}

According to a recent report, X-ray irradiation into the organic 
Mott insulator increases electrical conductivity at room temperature (RT)\cite{1441}.
Whereas the detailed mechanism of such carrier delocalization has not 
been clarified, a possible scenario is a kind of carrier doping to the
electronic system due to local destruction of the charge balance
between donor and anion molecules.
Apart from attractive mechanism of controlling carrier density,
here we focus on the permanent improvement of conductivity
by X-ray irradiation:
it can be readily applied to fabrication of metallic structures
in the insulating crystal by means of spatially local irradiation,
such as through mask pattern (Fig. \ref{f_schematic}(a)).
We employ this method as a first step to exploit a new tool for 
molecular electronic device fabrication.
In this letter, we fabricate a metallic structure in the Mott insulator
$\kappa$-(BEDT-TTF)$_2$Cu[N(CN)$_2$]Cl by X-ray irradiation through a
mesh mask.
Scanning microregion infrared reflectance spectroscopy (SMIS) \cite{1308}
provides imaging of the metallic domains with approximately
90 $\times$ 90 $\mu$m$^2$ area.
There is no visible irradiation damage under optical microscope and 
scanning tunneling microscope (STM).

\begin{figure}
\begin{center}
\includegraphics[clip,width=8cm]{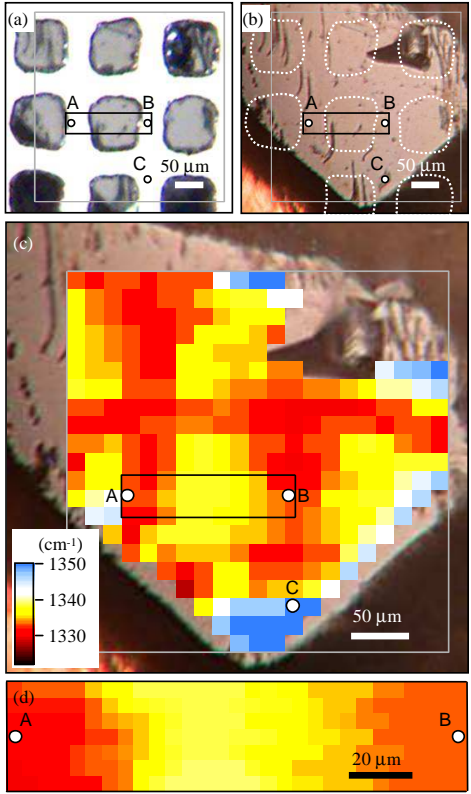}
\end{center}
\caption{
(a) Top view of the molybdenum mesh mask on the sample.
Two squares (gray and black) corresponds to the area scanned by the SMIS measurement.
(b) Crystalline facet of  $\kappa$-(BEDT-TTF)$_2$Cu[N(CN)$_2$]Cl after irradiation.
The mask positions are depicted with dotted curves.
(c) Peak frequency map of an intramolecular vibration mode ($\nu_3$)
measured at 4 K.
The IR spectra are taken at regular intervals of 15 $\mu$m.
(d) Fine scanning map in the area identical to the black squares shown in
Figs. \ref{f_map}(a)--(c). The spatial interval of the data is 5 $\mu$m (see Fig. \ref{f_R}(b)).
}
\label{f_map}
\end{figure}

\begin{figure}
\begin{center}
\includegraphics[clip,width=8cm]{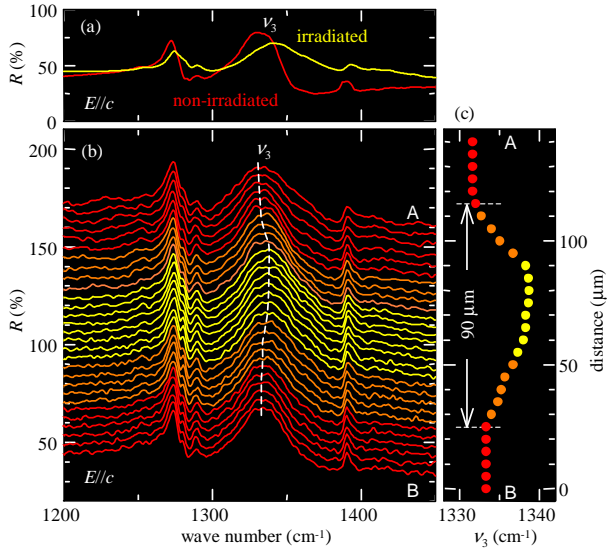}
\end{center}
\caption{
(a) Reflectance spectra of a non-irradiated
$\kappa$-(BEDT-TTF)$_2$Cu[N(CN)$_2$]Cl (yellow curve) and 
an irradiated one for 325 hours (red curve).
(b) Positional variation of the reflectance spectra of 
$\kappa$-(BEDT-TTF)$_2$Cu[N(CN)$_2$]Cl irradiated through the mesh mask.
The locations of the spectra correspond to the points A to B shown in 
Fig. \ref{f_map}. 
(c) Spatial dependence of the $\nu_3$ peak frequency for the spectra
shown in Fig. \ref{f_R}(b).
}
\label{f_R}
\end{figure}

Single crystals of $\kappa$-(BEDT-TTF)$_2$Cu[N(CN)$_2$]Cl were grown by a standard 
electrochemical technique.
The configuration for X-ray irradiation is schematically depicted in Fig. \ref{f_schematic}(a).
The largest facet corresponding to the conduction layer in an as-grown 
crystal with dimensions approximately 0.4$\times$0.3$\times$0.1 mm$^3$
was covered with a mask of molybdenum mesh sheet (with $\sim$90$\times$90 
$\mu$m$^2$ mesh size, $\sim$30 $\mu$m bar width, and 10 $\mu$m thickness, see Fig. \ref{f_map} (a)).
White X-ray (40 kV, 20 mA tungsten target) was irradiated through the 
mask for 160 hours at RT.
No other treatment on the sample surface was carried out.
Figure \ref{f_map}(b) is a photograph taken after irradiation, indicating that
there is no visible irradiation damage on the surface, while there are 
several inherent scratches and a large dent at the upper-right side.
We performed an SMIS study\cite{1308} using a synchrotron radiation
light source at SPring-8 BL43IR\cite{SPring8}.
Each polarized reflectance spectrum was measured on the irradiated plane at 4 K
with the electric field ($E$) parallel to the $c$-axis.
Whether or not a $\kappa$-(BEDT-TTF)$_2X$ ($X$ denotes anions) is metallic 
is indicated by a totally symmetric intramolecular 
vibration mode $\nu_3(a_g)$ \cite{1115} of the spectra.
Since this vibration mode markedly reflects the electronic state
of the conduction carrier via electron-molecular vibration coupling,\cite{1235}
the frequency of the $\nu_3$ mode is sensitive to the degree of
localization of the carriers; the peak frequency in a metallic
state is found to be higher than that in an insulating one \cite{1115}.

First of all, to verify this frequency shift in the present case, let us 
compare the polarized reflectance spectrum of a non-irradiated sample
with that of an irradiated one for 325 hours.
Figure \ref{f_R}(a) shows the reflectance spectra at 4 K with $E \parallel c$.
The most remarkable peak, found at approximately 1330 and 1345 cm$^{-1}$ in
the non-irradiated (red curve) and irradiated (yellow) samples,
respectively, comes from the $\nu_3$ phonon mode as mentioned above.
The peak shift toward higher frequency by irradiation clearly demonstrates that X-ray 
irradiation induces the metallic state.\cite{comment1}
This is in good agreement with the dc resistivity data \cite{1441}.

Figure \ref{f_R}(b) shows a spatial variation of the spectra taken at regular 
intervals of 5 $\mu$m from point A to B (denoted in Fig. \ref{f_map}).
Each spectrum is shifted by 4\% for clarity.
The peak frequency varies with position, reflecting 
higher frequency of the metallic sites (yellow curves) than that of the 
insulating sites (red curves).
The peak positions of the $\nu_3$ mode represented with a broken
curve are located between 1330 and 1340 cm$^{-1}$, which are
suitably consistent with the data displayed in Fig. \ref{f_R}(a).

Figure \ref{f_map} (c) shows the two-dimensional map of the peak frequency of the 
$\nu_3$ mode over 300$\times$300 $\mu$m$^2$ area.
The spectra were taken at regular intervals of 15 $\mu$m.
Since brighter color represents higher frequency,
the $\nu_3$ mode in the yellow-to-blue colored area
($\sim$1335 -- 1350 cm$^{-1}$) indicates the metallic state caused by
X-ray irradiation. 
By contrast, the red area with a frequency of approximately 1330 
cm$^{-1}$ expresses the insulating state featuring non-irradiation.
Thus the obtained image successfully describes the periodic metallic 
domains irradiated through the mask pattern shown in Fig. \ref{f_map}(a).

We next discuss the horizontal resolution of the domain edge.
The spatial dependence of the $\nu_3$ peak frequency between points A and 
B is shown in Fig. \ref{f_R}(c). 
This indicates that the metallic region marked with the orange and yellow
circles spreads over a distance of 90 $\mu$m, which coincides with
the mesh size of the mask.
The border between metallic and insulating regions (orange circles) is
in a range of $\sim$ 25 -- 30 $\mu$m.
This broad interface of the metallic domain is possibly originating
from diffusion of X-ray at the mask edge.
Moreover, we note that the area near the sample edge has relatively high
peak frequency (white-to-blue area in Fig. \ref{f_map} (c)).
There even exists some metallic sites regardless of its mask, for
example at around point C.
This out of alignment in the metallic pattern from the mask
may be caused by indirect exposure of backscattering from the substrate.
An effective method to improve sharpness of the boundary is to adopt
other mask materials which can more effectively absorb
irradiation, such as Pb or W.
We therefore prepared another irradiated sample masked with a Pb plate (with
$\sim$ 10 $\mu$m thickness) on half of the sample surface.
The similar SMIS measurement reveals that 
this yields sharper edge with a border of 10 -- 20 $\mu$m 
(not shown in Figures), which is comparable to the resolution limit of 
the present SMIS measurement.

It is noteworthy to comment on the vertical structure of the metallic domain.
Since the penetration depth of the incident infrared light in the SMIS 
measurement is roughly estimated as approximately 1 $\mu$m,\cite{1320} 
the metallic domain will have a depth of 1 $\mu$m at least.
Taking into account the high transmittance of X-ray, this bulky
modification is reasonably accepted.
However, decay of X-ray by distance may result in gradual exposure in
depth direction.
An extra SMIS study on the back side of the irradiated surface
would give a conclusive answer to the spatial blurring in depth.

\begin{figure}
\begin{center}
\includegraphics[clip,width=8cm]{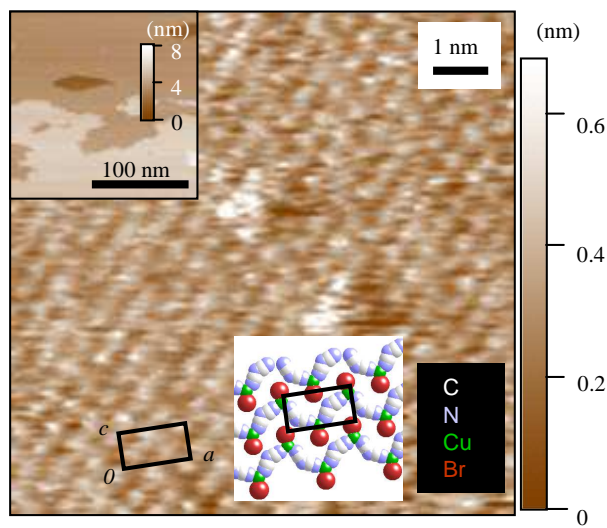}
\end{center}
\caption{
Topographic STM image of $\kappa$-(BEDT-TTF)$_2$ Cu[N(CN)$_2$]Br
irradiated for 160 hours.
As a possible assignment, a top view of the anion layer structure is 
overlaid on the image.
Although one can see two bright spots in the middle of the image,
such anomalous structures are frequently found in non-irradiated salts 
as well. The inset shows a wide area image.
}
\label{f_STM}
\end{figure}

To confirm no deterioration by irradiation, an STM measurement was 
performed in a range of nanometer to micrometer scales.
The RT-STM was carried out for an 
analogue sample $\kappa$-(BEDT-TTF)$_2$Cu[N(CN)$_2$]Br after 160 hours
irradiation. 
As shown in the inset of Fig. \ref{f_STM},
wide plateaus separated by step of approximately $\sim$ 4 nm are observed.
This terrace structure is widely recognized in many CT salts.
Moreover, the main panel of Fig. \ref{f_STM} shows a periodic pattern at 
nanometer scale, which is well consistent with the lattice constants.
It should be noted that there is no remarkable difference between the 
STM images in the irradiated and non-irradiated samples within the present
resolution.\cite{comment2}
An important finding from a viewpoint of application is that 
X-ray irradiation hardly destructs 
the crystal structure, implying that mechanical stability of the 
irradiated crystal differs little from that of the non-irradiated one.

Several practical applications of the present method can be proposed:
drawing the electrical circuit by X-ray microbeam with 
small spot and high directivity (Fig. \ref{f_schematic}(b)), a metallic dot 
formation inside the insulating crystal by convergent projection of 
X-ray beam (Fig. \ref{f_schematic}(c)),
and reforming mechanically and electronically stable interface between 
metal electrodes and organic crystal surface. 


In conclusion, we have fabricated the metallic pattern in the
Mott insulating crystal $\kappa$-(BEDT-TTF)$_2$Cu[N(CN)$_2$]Cl by means of local
X-ray irradiation.
The metallic domain of approximately 90$\times$90 $\mu$m$^2$
with a border width of $\sim$ 25 -- 30 $\mu$m was produced,
which was successfully visualized by the SMIS measurement.
No irradiation damage was observed in the STM measurement, implying
the mechanical stability of the irradiated region comparable to that of 
the non-irradiated one.
It guarantees that the present method is potentially applicable to fabricate
molecular electronic devices.

\acknowledgement
The authors thank T. Hirono and T. Kawase for their 
technical support.
Synchrotron radiation measurements were performed at SPring-8, BL43IR with the 
approval of JASRI (2007B1050 and 2008A1121).
The research was supported by a Grant-in-Aid for Scientific Research 
(Nos. 16076201, 17340099, and 18654056) from the Ministry of Education, 
Culture, Sports, Science and Technology and the Japan Society for the 
Promotion of Science.


\begin{thebibliography}{9}
\bibitem{comment1}
In fact, the infrared spectrum of the irradiated sample exhibits an 
appearance of Drude-like structure characteristic of itinerant carriers, 
which will be reported elsewhere.
\end{thebibliography}
\end{document}